\documentclass[a4paper]{article}
\usepackage{microtype}
\usepackage{url}
\usepackage{INTERSPEECH2022}
\usepackage{amsmath}
\usepackage[linesnumbered]{algorithm2e}
\usepackage{algpseudocode}
\usepackage{bm}
\usepackage{adjustbox}
\usepackage{caption}
\RestyleAlgo{ruled} 

\usepackage{amsmath,amsfonts,bm}

\def\Figref#1{Figure~\ref{#1}}
\def\Tabref#1{Table~\ref{#1}}

\def\Secref#1{Section~\ref{#1}}

\def\eqref#1{equation~\ref{#1}}
\def\Eqref#1{Equation~\ref{#1}}

\def\1{\bm{1}}

\def\eps{{\epsilon}}

\def\vx{{\mathbf{x}}}
\def\vy{{\mathbf{y}}}

\def\evx{{x}}
\def\evy{{y}}

\def\mM{{\mathbf{M}}}

\def\mW{{\bm{W}}}
\def\mX{{\mathbf{X}}}
\def\mY{{\mathbf{Y}}}

\DeclareMathAlphabet{\mathsfit}{\encodingdefault}{\sfdefault}{m}{sl}
\SetMathAlphabet{\mathsfit}{bold}{\encodingdefault}{\sfdefault}{bx}{n}

\usepackage{booktabs}
\usepackage[table,xcdraw]{xcolor}
\usepackage{color}
\usepackage{soul}
\usepackage{siunitx}

\usepackage[colorlinks,linkcolor=blue,]{hyperref}
 
\title{Neural Vocoder is All You Need for Speech Super-resolution}
\name{
Haohe Liu$^{1}$,
Woosung Choi$^2$,
Xubo Liu$^{1}$, 
Qiuqiang Kong$^3$, 
Qiao Tian$^3$, 
DeLiang Wang$^4$
}

\address{
  $^1$Centre for Vision, Speech and Signal Processing (CVSSP), University of Surrey, UK\\
  $^2$Centre for Digital Music, Queen Mary University of London, UK\\
  $^3$Speech, Audio, and Music Intelligence (SAMI) Group, ByteDance, China\\
  $^4$Department of Computer Science and Engineering, The Ohio State University, USA
  }
\email{hl01486@surrey.ac.uk,  wschoi.cs@gmail.com, dwang@cse.ohio-state.edu}

\begin{document}

\maketitle
%
\begin{abstract}
Speech super-resolution~(SR) is a task to increase speech sampling rate by generating high-frequency components. Existing speech SR methods are trained in constrained experimental settings, such as a fixed upsampling ratio. These strong constraints can potentially lead to poor generalization ability in mismatched real-world cases. In this paper, we propose a \textbf{n}eural \textbf{v}ocoder based speech \textbf{s}uper-\textbf{r}esolution method~(NVSR) that can handle a variety of input resolution and upsampling ratios. NVSR consists of a mel-bandwidth extension module, a neural vocoder module, and a post-processing module. Our proposed system achieves state-of-the-art results on VCTK multi-speaker benchmark. On \num{44.1} kHz target resolution, NVSR outperforms WSRGlow and Nu-wave by \num{8}\% and \num{37}\% respectively on log-spectral-distance and achieves a significantly better perceptual quality. We also demonstrate that prior knowledge in the pre-trained vocoder is crucial for speech SR by performing mel-bandwidth extension with a simple replication-padding method. Samples can be found in {\small \url{https://haoheliu.github.io/nvsr}}.

\end{abstract}


\noindent\textbf{Index Terms}: neural vocoder, speech super resolution, bandwidth extension, deep learning, flexible resolution

\section{Introduction}
Speech super-resolution (SR) aims to improve speech fidelity by predicting the high-resolution~(HR) speech given the low-resolution~(LR) speech. Low-resolution speech is common for reasons such as compression and low sampling rate. From a spectrogram point of view, speech SR is also equally referred to as bandwidth extension~(BWE)~\cite{BWE-ekstrand2002bandwidth}. SR is an important technique for real-world applications, including speech quality enhancement~\cite{chennoukh2001speech} and text-to-speech synthesis~\cite{nakamura2014mel}. 

A lot of early studies ~\cite{nakatoh2002generation-bwe-linear-mapping,kontio2007neural-network-bwe-2} break SR into spectral envelop estimation and excitation generation from LR. At that time, the direct mapping from LR to HR is not widely explored since the dimension of HR is relatively high. Later, SR methods based on deep neural network~\cite{li2015dnn-bwe,audio-supre-resolution-SR-kuleshov2017audio,gupta2019wavenet-bwe,eskimez2019speech} show better subjective quality than traditional methods. Most neural network based methods study the \num{8} kHz to \num{16} kHz or \num{4} kHz to \num{16} kHz sampling rate upsample problems, such as AECNN~\cite{heming-towards-sr-wang2021towards} and TFNet~\cite{tf-network-sr-lim2018time}. Recently, Nu-wave~\cite{nu-wave-lee2021nu} and WSRGlow~\cite{zhang2021wsrglow} have explored the higher \num{48} kHz target resolution, where the input sampling rate is usually \num{12} kHz, \num{16} kHz, or \num{24} kHz. 

Existing speech SR studies are usually performed under controlled experimental settings. For example, the input resolution and bandwidth in data simulations is always fixed value during training and evaluation.
However, in real-world scenarios, the speech SR applications generally require the capability of handling diverse settings such as different input resolutions and bandwidths. 
These mismatches potentially lead to degraded speech SR performance. Also, the demand for high upsampling-ratio~(UPR) SR is common on low-quality historical recordings, where UPR stands for target bandwidth divided by input bandwidth. Nevertheless, less attention has been given to the high UPR cases. High UPR speech SR is challenging because it has more low-frequency harmonic structures to be predicted. To our knowledge, most speech SR works only experiment on the UPR around \num{3}, \num{4}, and \num{6}~\cite{nu-wave-lee2021nu, zhang2021wsrglow, kim2021learning}.

To alleviate the difficulties in addressing flexible input resolution and high UPR mentioned above, we propose a neural-vocoder-based approach~(NVSR) for speech SR. We train NVSR in two separate stages: (1) an HR mel spectrogram prediction stage; (2) a vocoder waveform synthesis and post-processing stage. Instead of predicting HR in the linear frequency scale as shown in previous studies~\cite{tf-network-sr-lim2018time,hu2020phase}, HR prediction on a low-dimensional mel frequency scale is more tractable, especially in high UPR cases. Neural Vocoder~\cite{hifi-gan-kong2020hifi, tian2020tfgan} is the model that upsample a low dimensional feature~(e.g., mel spectrogram) to high-resolution waveforms. This coarse to fine generation process is similar to speech SR. Thus the prior knowledge of speech provided by the neural vocoder, such as phoneme structures, would be useful for speech SR. 

To our knowledge, we are the first attempt that tackles the flexible input resolution and high UPR problems for speech SR. Our proposed NVSR can achieve a UPR up to \num{22.05} (from \num{2} kHz to \num{44.1} kHz).
NVSR achieves state-of-the-art result on the VCTK Multi-speaker benchmark under multiple experimental settings, outperforming previous methods by a large margin. We observe that even replacing neural network methods with a simple replication-padding method without learning, NVSR can still significantly outperform existing speech SR methods. This demonstrates the superior performance offered by the neural vocoder for speech SR. Unlike most previous studies, we mainly focus on the multi-speaker setup, which is more challenging than the single-speaker setting studied in other literature~\cite{heming-towards-sr-wang2021towards}. Our code and pre-trained models are open-sourced~\footnote{ \url{https://github.com/haoheliu/ssr\_eval}} to facilitate reproducibility.

The paper is organized as follows. \Secref{sec:problem-formulation} introduces the problem formulation of speech SR. \Secref{sec:methodology} introduces the pipeline of NVSR.
\Secref{sec:experiment} discusses the experimental settings and analysis the results. ~\Secref{sec:conclusion} draw a final conclusion.

\section{Problem Formulation}
\label{sec:problem-formulation}
Given a discrete signal $\vx_{l}=[\evx_{i}]_{i=1,2,...T\cdot l}$ sampled at sampling rate $l$, speech SR system estimates signal $\vy_{h}=[\evy_{i}]_{i=1,2,...T\cdot h}$, where $T$ is the length in seconds and $h>l$. According to the Nyquist theory, the highest bandwidth of $\vx_{l}$ and $\vy_{h}$ are $l/2$~(Hz) and $h/2$~(Hz), respectively. So, $\vx_{l}$ does not contain the frequency information between $h/2-l/2$~(Hz), which can be considered as the generation target of the SR task. 

Most frequency domain methods~\cite{audio-supre-resolution-SR-kuleshov2017audio, tf-network-sr-lim2018time} perform SR by predicting HR spectorgram from LR spectrogram, and transformed to waveform using the HR spectrogram prediction. This process can be formulated as 
\begin{equation}
  \label{eq:basic-equation}
  \vy_{h}=F^{-1}(G(F(\vx_{h}))),
\end{equation}
where $F$ and $F^{-1}$ are the time to frequency spectrogram transformation and it's reverse transformation respectively. $\vx_{h}$ is the upsampled version of $\vx_{l}$, and function $G(\cdot)$ takes band-limited spectrogram as input and output full band spectrogram. Here $\vx_{l}$ is upsampled to ensure the system input and output have the same shape and the model design is not restricted by the ratio $h/l$. In most studies, $F$ and $F^{-1}$ are short-time-fourier-transform~(STFT) and inverse STFT~(iSTFT), respectively.



\section{Methodology}
\label{sec:methodology}
\begin{figure}[htbp] 
  \centering
  \includegraphics[page=1,width=0.99\linewidth]{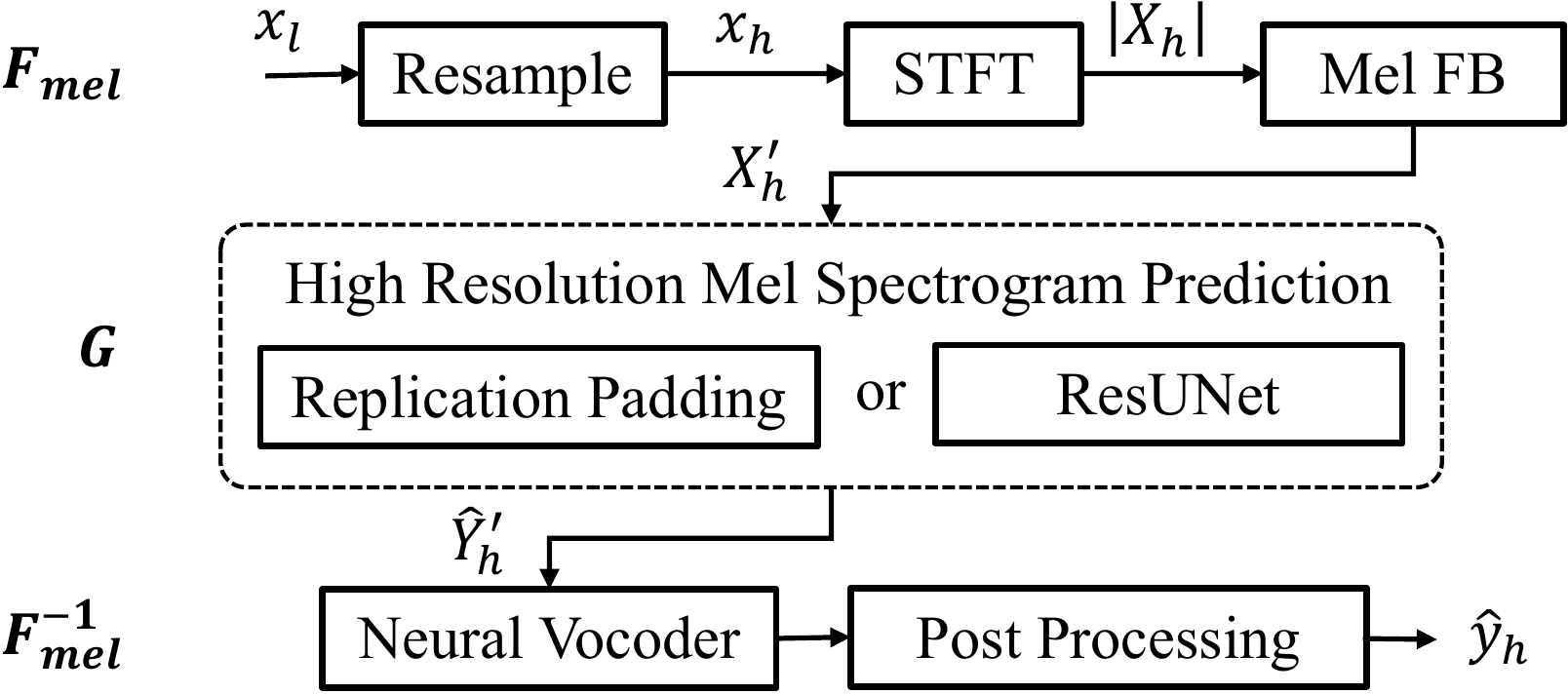}
  \caption{Overview of the NVSR pipline.}
  \label{fig-main}
\end{figure}

\Figref{fig-main} shows the overall pipline of NVSR including mel transform $F_\text{mel}$, high resolution mel spectrogram prediction with $G$, and inverse mel transform $F_\text{mel}^{-1}$. Similar to \Eqref{eq:basic-equation}, this process can be written as
\begin{equation}
  \label{eq: NVSR_equation}
  \hat{\vy}_{h}=F_\text{mel}^{-1}(G(F_\text{mel}(\vx_{h}))) \approx V_{\phi}(G(F_\text{mel}(\vx_{h}))).
\end{equation}
Here $G$ aims to predict the HR mel spectrogram with the LR input, as shown in~\Eqref{eq:spec-main-idea}. 
\begin{equation}
  \label{eq:spec-main-idea}
  G: \mX_\text{mel} \to \mY_\text{mel},
\end{equation}
where $\mX_\text{mel}$ and $\mY_\text{mel}$ stands for the mel spectrogram of $\vx$ and $\vy$. $\mX_\text{mel}$ is calculated by $F_\text{mel}(\vx)=\left| \mX \right| \mW$, in which $\mX$ is the STFT of $\vx$ and $\mW$ is a set of mel filter banks.

Note that the inverse mel transform is not mathematically solvable. Thus, $F_\text{mel}^{-1}$ in NVSR is modeled by a neural vocoder $V_{\phi}$. We show that $V_{\phi}$ achieves almost the same perceptual quality as the ideal inverse transform. For mel spectrogram prediction, we first introduce a ResUNet based method. Then we introduce a non-parametric replication-padding method, which mainly exploits the prior knowledge learned by the vocoder. 

\subsection{Mel Spectrogram Prediction}
\label{sec:mel spectrogramtrogram-prediction}

\subsubsection{ResUNet-based Method}
\label{sec:nn-method}

\begin{figure}[tbp] 
  \centering
  \vspace{-1.6em}
  \includegraphics[page=1,width=0.90\linewidth]{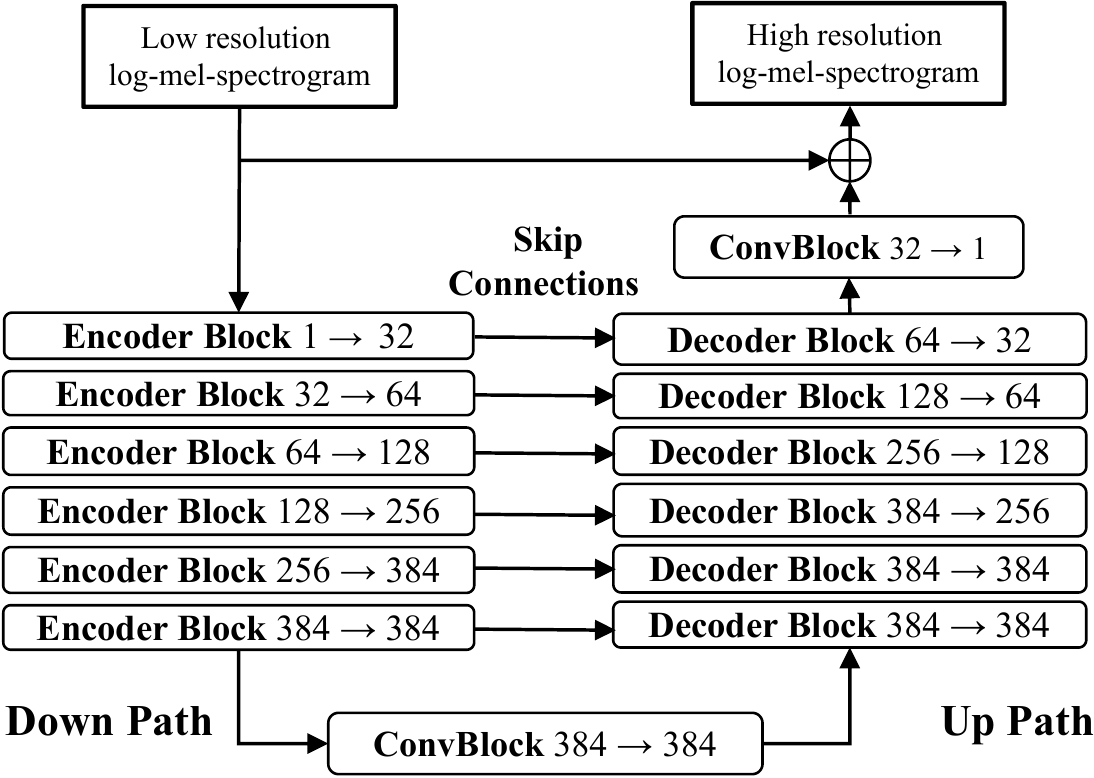}
  \caption{Predicting the higher frequnecies of mel spectrogram with ResUNet. The numbers in each block means input and output channels.}
  \label{fig-resunet-method}
\end{figure}

We use ResUNet~\cite{kong2021decoupling, liu2021cws} to model the $G(\cdot)$ in \Eqref{eq:spec-main-idea} and estimate $\mY_\text{mel}$. This process can be written as
\begin{equation}
    \label{eq:unet}
    \hat{\mY}_\text{mel} = \text{ResUNet}(\mX_\text{mel}) \odot (\mX_\text{mel} + \eps),
\end{equation}
where $\odot$ means element-wise multiplication, and $\eps$ is a small constant to avoid the zero values in $\mX_\text{mel}$. As shown in~\Figref{fig-resunet-method}, ResUNet consists of six encoder and six decoder blocks. There are skip connections between encoder and decoder blocks at the same level. Both encoder and decoder blocks share the same structure, which is four convolutional blocks~(ConvBlock). Each ConvBlock consists of two convolution operations with batch normalization~\cite{ioffe2015batch}, and leakyReLU~\cite{Maas13rectifiernonlinearities} activation. The encoder blocks apply average pooling for downsampling. The decoder blocks apply transpose convolution for upsampling. After the last decoder block, we apply one more ConvBlock to estimate the output residual, which is added with the input to form the final estimation. Note that we use addition in~\Figref{fig-resunet-method} instead of multiply in~\Eqref{eq:unet} because we use log scale during our implementation.
ResUNet is optimized using the mean absolute error~(MAE) loss between the estimated mel spectrogram $\hat{\mY}_\text{mel}$ and the target mel spectrogram $\mY_\text{mel}$.

\begin{equation}
  \label{eq:neural-network-based-loss}
  loss = \frac{1}{T F}\sum^{T}\sum^{F}|\mY_\text{mel}-\hat{\mY}_\text{mel}|
\end{equation}

\subsubsection{Replication padding-based Method}
\label{sec:statical-method}

Replication padding-based methods do not use any training data for mel spectrogram prediction. Instead, it directly copies the energy of the cutoff frequency to the higher frequency bands. We use this method to demonstrate the importance of the neural vocoder. 
Specifically, we first search the cutoff frequency $c$ based on the energy of $\mX_\text{mel}$. Then we construct the frequency cutoff mask $\mM_{T\times F}$ with binary values, where the frequency indexs greater than $c$ are all zeros and others are all ones. $\mM_{T\times F}$ is used later to select the higher/lower frequencies bands. 
Finally, we repeat the energy of cutoff frequency for $F-c$ times and add it with LR mel spectrogram. The output is the HR mel spectrogram estimation $\hat{\mY}_\text{mel}$. This process can be written as~\Eqref{eq:padding}
\begin{equation}
\label{eq:padding}
\hat{\mY}_\text{mel} = \mM \odot \mX_\text{mel} + |1-\mM| \odot  \left( \mX_\text{mel}[:,c] \cdot \textbf{1}_{1\times F} \right),
\end{equation}
where $\textbf{1}_{1\times F}$ is the all-one matrix, and $\mX_\text{mel}[:,c]$ is the energy distribution across time at the cutoff frequency $c$.


\subsection{Neural Vocoder}

We choose TFGAN~\cite{tian2020tfgan} as the neural vocoder $V_{\phi}$. TFGAN can directly upsample mel spectrogram into waveform with transpose convolution and one-dimensional convolutional neural networks. By incorporating multi-resolution losses and discriminators in both the time and frequency domains, TFGAN can achieve state-of-the-art performance on vocoding. In order to achieve speaker-independent, TFGAN is trained with a large corpus of more than one thousand speakers. As reported in ~\cite{liu2021voicefixer}, the mean opinion score~(MOS) of the open-sourced TFGAN\footnote{\url{https://github.com/haoheliu/voicefixer}} is \num{3.74}, with the ground truth MOS score at \num{3.95}.

\subsection{Post Processing} 

\begin{figure}[htbp] 
  \centering
  \includegraphics[page=5,width=0.99\linewidth]{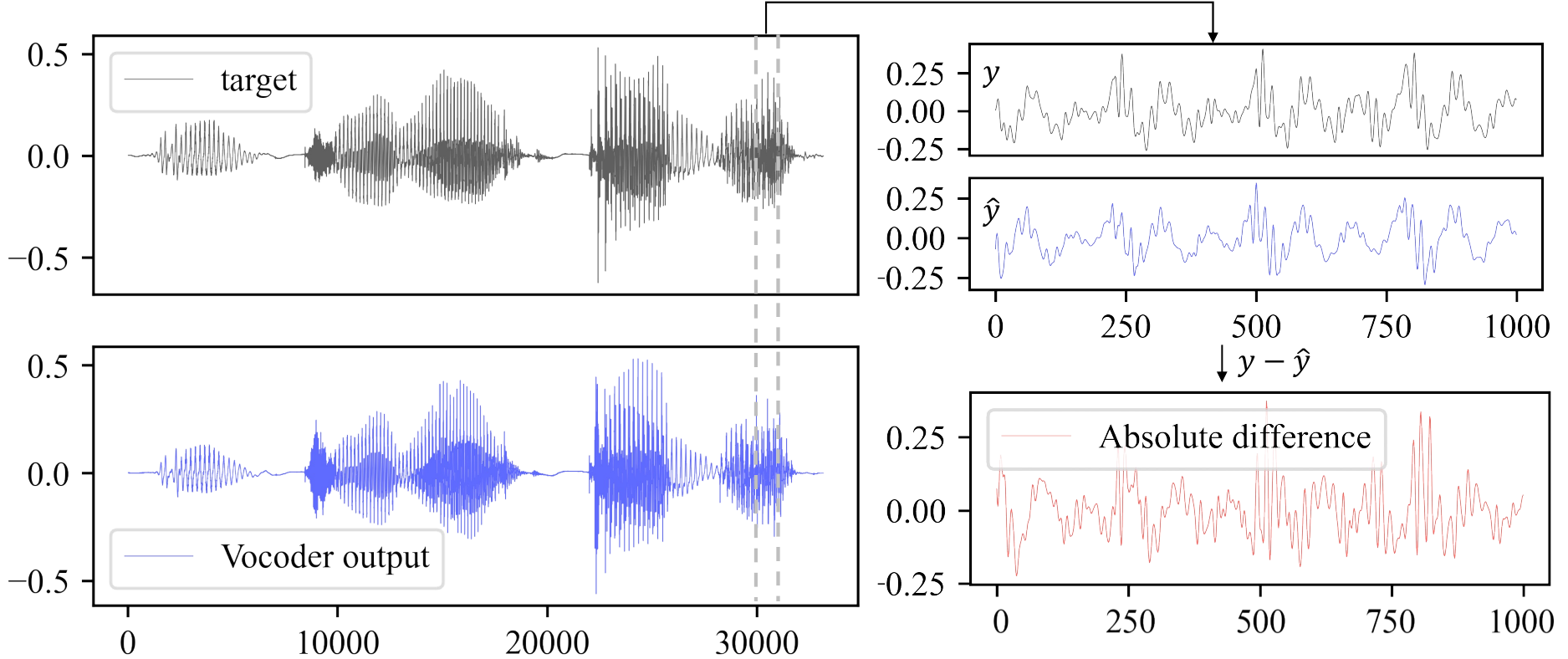}
  \caption{Visualization of the difference between the vocoder output and target waveforms.}
  \label{fig-postprocess-method}
\end{figure}

Even though the output of TFGAN can achieve good perceptual quality, its output still has trivial differences from the target, which makes it achieve poor results in the objective evaluation. For example, as shown in \Figref{fig-postprocess-method}, the vocoder output looks the same as the target in the waveform, but their absolute difference is still substantial. This is a common evaluation problem in generative model~\cite{nu-gan-kumar2020nu, liu2019speech, liu2021conditional}. To alleviate the problem in evaluation, we propose to perform a lower-frequencies replacement~(LFR) operation on the vocoder output as post-processing. Since the original lower-frequencies information in the input usually does not need to change, we replace the low-frequency bands in the vocoder output with the original input. Experimental result shows this can improve the metrics score.

\section{Experiment}
\label{sec:experiment}
\subsection{Dataset}
We build the train and test sets using VCTK~(version 0.92)~\cite{vctk-yamagishi2019cstr}, a multi-speaker English corpus that contains \num{110} speakers with different accents. We split it into a part for train~(VCTK-Train) and a part for test~(VCTK-Test). Following the data preparation strategy of~\cite{nu-wave-lee2021nu}, only the \textit{mic1} microphone data is used for experiments, and \textit{p280} and \textit{p315} are omitted for the technical issues. For the remaining 108 speakers, the last eight speakers\footnote{p360, p361, p362, p363, p364, p374, p376, s5} constitute the VCTK-Test. The remaining \num{100} speakers are defined as VCTK-Train. For the training of NVSR, all the utterances are resampled at the \num{44.1} kHz sample rate. 

We follow the LR simulation process in ~\cite{audio-supre-resolution-SR-kuleshov2017audio,heming-towards-sr-wang2021towards}.
Given a target speech $\vy_{h}$ in VCTK-Train, to obtain the low-resolution $\vx_{l}$, we first convolve $\vy_{h}$ with an order eight \textit{Chebyshev type I} lowpass filter with cutoff frequency $l/2$. Then we subsample the signal to $l$ sample rate using polyphase filtering. We evaluate the performance of each system on different low sampling rates $l$.

\subsection{Evaluation Metrics}

We use Log-spectral distance~(LSD) as the evaluation metrics following~\cite{heming-towards-sr-wang2021towards, nu-wave-lee2021nu, nu-gan-kumar2020nu}. For target signal $\vy_{h}$ and output estimation $\hat{\vy}_{h}$, LSD can be computed as \Eqref{metrics-lsd}, where $\mY$ and $\hat{\mY}$ stand for the magnitude spectrogram of $\vy_{h}$ and $\hat{\vy}_{h}$. A lower LSD value indicates better SR performance. We report the mean LSD of the VCTK-Test as the final score of a system. 
\begin{equation}
    \label{metrics-lsd}
    \vspace{-1.0em}
    \mathrm{LSD}(\mY,\hat{\mY}) = \frac{1}{T}\Sigma_{t=1}^{T}\sqrt{\frac{1}{F}\Sigma_{f=1}^{F}\log_{10}(\frac{\mY(f,t)^2}{\hat{\mY}(f,t)^2})^2}
\end{equation}

\subsection{Baselines}
We include several state-of-the-art methods as baselines. Since NVSR uses a target sampling rate at 44.1 kHz, the baseline with 48 kHz output is downsampled to 44.1 kHz for a fair comparison. Similarly, the output of NVSR can also be downsampled to compare with other baselines with a smaller target sampling rate~(e.g., 16 kHz). For the 48 kHz SR model, we reproduced Nu-wave~\footnote{\url{https://github.com/mindslab-ai/nuwave}} and WSRGlow~\footnote{\url{https://github.com/zkx06111/WSRGlow}} with their open-sourced code and default settings. We train each Nu-wave for \num{200} epochs and each WSRGlow for 100k iterations following~\cite{zhang2021wsrglow}. For \num{16} kHz target sampling rate, we compared NVSR with the result reported in AECNN~\cite{heming-towards-sr-wang2021towards}, TFNet~\cite{tf-network-sr-lim2018time}, and AudioUNet~\cite{audio-supre-resolution-SR-kuleshov2017audio}.

\subsection{Training Details}
The training high-resolution and low-resolution data pairs are built on the fly. we uniformly randomize the cutoff frequency ${l/2}$ of the training data in [\num{1}, \num{16}] kHz. We use an Adam optimizer~\cite{kingma2014adam} with $\beta_1=0.5,\beta_2=0.999$ and a \num{3e-4} learning rate to optimize the ResUNet. We apply the first \num{1000} steps as the warmup phase, during which the learning rate grows linearly from \num{0} to \num{3e-4}. Then the learning rate is decayed by \num{0.85} every epoch. We stop the training after \num{60} epochs. For all the STFT and iSTFT, we use the Hanning window with a window length of \num{2048} and a hop length of \num{441}. We use \num{128} mel filterbanks to calculate the mel spectrogram. We use eight Nvidia-V100-32GB GPUs to train the ResUNet, which takes about \num{3.9} hours.

\subsection{Result and Discussion}
\label{sec:result}
\begin{table}[tbp]
\centering
\caption{LSD on VCTK-Test with 44.1 kHz target sampling rate. Model with $^{*}$ means it is trained with a fixed input resolution.}
\label{tab:48k}
\begin{tabular}{@{}lcccccl@{}}
\toprule
Input sampling rate~(kHz)   & 4 & 8 & 12 & 24  \\ \midrule
$^{*}$Nu-wave~\cite{nu-wave-lee2021nu}~(3.0M$\times 4$) &  1.42  &  1.42  &  1.40 & 1.22     \\
$^{*}$WSRGlow~\cite{zhang2021wsrglow}~(229.9M$\times 4$) & 1.12  & 0.98 & 0.87  & 0.79    \\
NVSR-Pad~(33.9M)    & 1.54 & 1.46 & 1.18 & 0.91    \\
NVSR~(99.0M)    & \textbf{0.98} & \textbf{0.91} & \textbf{0.85} & \textbf{0.70}    \\ \bottomrule
\end{tabular}
\end{table}

\begin{table}[tbp]
\centering
\caption{LSD on VCTK-Test with 16 kHz target sampling rate.}
\label{tab:16k}
\begin{tabular}{@{}lcccc@{}}
\toprule
          Input sampling rate~(kHz)                     & 2   & 4  & 8   \\ \midrule
{ $^{*}$AudioUNet~\cite{audio-supre-resolution-SR-kuleshov2017audio}~(70.9M$\times 3$)} & N/A  & 1.40 &  1.32  \\
{ $^{*}$TFNet~\cite{tf-network-sr-lim2018time}~(58.8M$\times 3$)}    & N/A & N/A   & 1.36   \\
{ $^{*}$AECNN~\cite{heming-towards-sr-wang2021towards}~(10.2M$\times 3$)}    & N/A  & \textbf{0.95} & 0.88   \\
{NVSR-Pad~(33.9M)}                             &  2.51 & 2.24 & 1.89    \\
{NVSR~(99.0M)}                             &  \textbf{1.07} & \textbf{0.95} & \textbf{0.78}   \\ \bottomrule
\end{tabular}
\vspace{-0.6em}
\end{table}

\begin{figure*}[tbp] 
  \centering
  \includegraphics[page=7,width=0.99\linewidth]{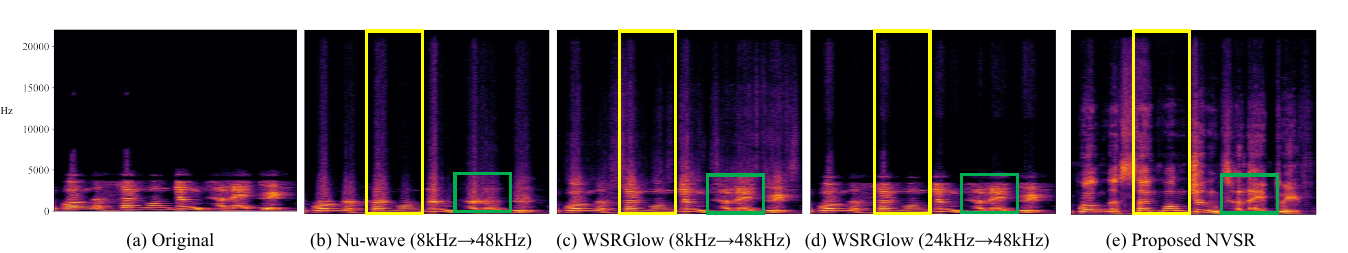}
  \caption{Robustness test with a low-resolution audio~(a) from an old movie\protect\footnotemark. No ground truth is available. }
  \label{fig-tiedaoyoujidui}
\end{figure*}
\footnotetext{\url{https://www.youtube.com/watch?v=A7Q7fmIuKdM}} 

\begin{table*}[htbp]
\begin{minipage}{0.65\textwidth}
\caption{Evaluation results on VCTK-Test with different input sampling rate settings. Note that NVSR can also work on other input sampling-rates between 2 kHz to 32 kHz.}
\label{table:main_result}
\begin{tabular}{@{}lcccccccc@{}}
\toprule
Input sampling rate~(kHz)                & 2 & 4 & 8 & 12 & 16 & 24 & 32 & AVG  \\ \midrule
Unprocessed          & 5.69 & 5.50 & 5.15 & 4.85  & 4.54  & 3.84  & 2.95  & 4.65 \\
Groundtruth-mel & 0.87 & 0.85 & 0.81 & 0.78  & 0.74  & 0.66  & 0.59  & 0.76 \\ \midrule
NVSR-Pad        & 1.55 & 1.54 & 1.46 & 1.18  & 1.11  & 0.91  & 0.76  & 1.21 \\
NVSR & \textbf{1.04} & \textbf{0.98} & \textbf{0.91} & \textbf{0.85}  & \textbf{0.79}  & \textbf{0.70}  & \textbf{0.60}  & \textbf{0.84} \\
\textit{\textbf{w/o} post proc.}       & 1.06 & 1.02 & 0.98 & 0.96  & 0.93  & 0.91  & 0.89  & 0.96 \\
\textit{\textbf{w/o} mel pred.}   & 2.37 & 2.37 & 2.16 & 1.95  & 1.77  & 1.48  & 1.13  & 1.89 \\ \bottomrule
\end{tabular}
\end{minipage}\hfill
\begin{minipage}{0.30\textwidth}
\includegraphics[page=9,width=0.99\linewidth]{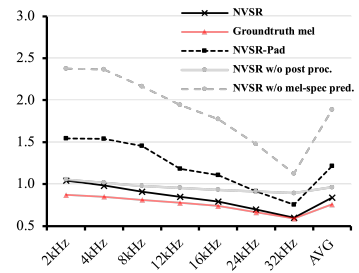}
\captionof{figure}{Visualization of Table.~\ref{table:main_result}}
\label{fig-visualization_of_table}
\end{minipage}
\vspace{-1em}
\end{table*}

\Tabref{tab:48k} and~\Tabref{tab:16k} compare NVSR with other state-of-the-art models on both 44.1 kHz and 16 kHz target sampling rates. We use $^{*}$ to denote a model is trained with a fixed input resolution. For example, we report four different sampling rate settings in~\Tabref{tab:48k} after training four different WSRGlows. Meanwhile, a single NVSR here can handle four different settings.

In \Tabref{tab:48k}, NVSR achieves an average LSD of \num{0.86}, outperforming WSRGlow's \num{0.94} and Nu-wave's \num{1.37} by \num{0.08} and \num{0.51}. The total parameter number of NVSR is \num{99.0} million~(M), of which \num{33.9}M comes from the vocoder. While WSRGlow is the largest model with \num{229.9}M parameters on each setting. This demonstrates the state-of-the-art performance and the efficiency of NVSR. NVSR-Pad is the replication padding-based NVSR. NVSR-Pad can already achieve an average LSD of \num{1.27}, outperforming Nu-wave on \num{12} kHz and \num{24} kHz. This proves the prior knowledge in vocoder can map the constant energy in the padded higher-frequency bands into meaningful energy distribution on the spectrogram. 
In \Tabref{tab:16k}, NVSR achieves the best performance on the \num{8} kHz input sampling rate. At \num{4} kHz, NVSR has a similar LSD score as AECNN. NVSR is the first to try \num{2} kHz to \num{16} kHz SR and achieve an LSD of \num{1.07}. Note that the NVSR in~\Tabref{tab:16k} and~\Tabref{tab:48k} is the same model.

In~\Figref{fig-tiedaoyoujidui}, we visualize the outputs of different methods on an old movie recording. The original speech is in low resolution, with the highest frequency around \num{4} kHz. As shown in the yellow boxes, the output of NVSR~(\Figref{fig-tiedaoyoujidui}e) contains properly shaped harmonic components, while other methods~(\Figref{fig-tiedaoyoujidui}b, and c) mainly fill in the high frequencies with stochastic components. WSRGlow trained with \num{24} kHz input resolution~(\Figref{fig-tiedaoyoujidui}d) fails to predict because of the mismatch between the input resolution~(\num{8} kHz) and the resolution of its training data~(\num{24} kHz). Meanwhile, we found NVSR is the only model that can repair the distribution in lower-frequency bands~(green boxes). Note that here we do not use post-processing in NVSR because the original input is of low quality. 


In \Tabref{table:main_result}, Groundtruth-mel stands for the system using ground truth mel spectrogram directly as the input to the vocoder. This equals the model performance when the mel spectrogram prediction module works flawlessly. This experiment marks the ideally best performance of the NVSR system, with an average LSD of \num{0.76}. We also tried to remove the post-processing LFR operation in NVSR, which degrade the average performance from \num{0.84} to \num{0.96}. If we do not perform mel spectrogram prediction on NVSR, the average LSD is \num{1.89}, which is still better than the unprocessed audio with a \num{4.65} LSD. This result means even without the mel prediction stage, the vocoder can still improve the metrics performance on the evaluation set.


To better understand the result, we visualize~\Tabref{table:main_result} in~\Figref{fig-visualization_of_table}. The red line is the theoretical best performance of our proposed system using pre-trained TFGAN. The result of NVSR is close to red line, meaning our mel spectrogram prediction module works well. \Figref{fig-visualization_of_table} also shows the performance of NVSR will degrade without post-processing, especially in high input sampling rate cases. On \num{24} kHz input sampling rate, the performance of the NVSR-Pad is even on par with the NVSR without post-processing. Note that the proposed post-processing operation can also be applied to other methods like Nu-wave and is very likely to increase the model performance on the test set. 

There are certain limitations of our proposed method. The performance of NVSR largely relies on the neural vocoder, which may become the bottleneck of NVSR. 
A possible future solution can be fine-tuning two stages in an end-to-end manner or designing an additional post-processing model for refinements. Besides, when extending NVSR to other kinds of sound, like music, vocoder may not as easy to train as the speech vocoder. 

\section{Conclusion}
\label{sec:conclusion}
This paper presents a novel and powerful neural vocoder-based speech super-resolution model NVSR. It shows strong performance across a wide range of input sampling rates between \num{2} kHz to \num{32} kHz. On the VCTK Multi-Speaker SR benchmark, NVSR outperforms the state-of-the-art models trained with different input resolution settings both on \num{16} kHz and \num{44.1} kHz evaluation sets. We also demonstrate that prior knowledge in vocoder is crucial to speech SR using a simple replication padding-based mel spectrogram prediction method. 

\bibliographystyle{IEEEtran}

\bibliography{mybib}

\end{document}